\documentclass[prl,showpacs,superscriptaddress,preprintnumbers,twocolumn,amsmath,nofootinbib]{revtex4}
 \usepackage[dvips,final]{graphicx}
  \usepackage{amssymb,amsmath,epsfig,bm,pifont}
   \graphicspath{{./figs/}}
\usepackage{color}

\usepackage{relsize}
\newcommand{\babar}{{\mbox{\slshape B\kern-0.1em{\smaller A}\kern-0.1em
            B\kern-0.1em{\smaller A\kern-0.2em R}}}
           }
\begin{document}
\thispagestyle{empty}
 \date{\today}
  \preprint{\hbox{RUB-TPII-02/2012}}
  \preprint{\hbox{FTUV/12-0516}}

\title{Comparing antithetic trends of data for the
       pion-photon transition form factor \\ }
 \author{A.~P.~Bakulev}
  \email{bakulev@theor.jinr.ru}
   \affiliation{Bogoliubov Laboratory of Theoretical Physics, JINR,
                141980 Dubna, Russia\\}

 \author{S.~V.~Mikhailov}
  \email{mikhs@theor.jinr.ru}
   \affiliation{Bogoliubov Laboratory of Theoretical Physics, JINR,
                141980 Dubna, Russia\\}

 \author{A.~V.~Pimikov}
  \email{alexandr.pimikov@uv.es}
   \affiliation{Bogoliubov Laboratory of Theoretical Physics, JINR,
                141980 Dubna, Russia\\}
   \affiliation{Departamento de F\'{\i}sica Te\'orica -IFIC,
                Universidad de Valencia-CSIC, E-46100 Burjassot
                (Valencia), Spain\\}

 \author{N.~G.~Stefanis}
  \email{stefanis@tp2.ruhr-uni-bochum.de}
   \affiliation{Institut f\"{u}r Theoretische Physik II,
                Ruhr-Universit\"{a}t Bochum,
                D-44780 Bochum, Germany}

\begin{abstract}
We perform a comparative theoretical study of the data at spacelike
momentum transfer for the
$\gamma^*\gamma\to\pi^0$
transition form factor, just reported by the Belle Collaboration,
vs. those published before by BaBar, also including the older CLEO
and CELLO data.
Various implications for the structure of the $\pi^0$ distribution
amplitude vis-\`a-vis those data are discussed and the existing
theoretical predictions are classified into three distinct categories.
We argue that the actual bifurcation of the data with antithetic trends
is artificial and reason that the Belle data are the better option.
\end{abstract}

\maketitle

The newly released data for the pion-photon transition form factor
(TFF)
$
 F^{\gamma^*\gamma\pi^0}(Q^2, q^2 \to 0)
$
by the Belle Collaboration \cite{Belle12}, seem to be dramatically
different from those reported in 2009 by the \babar Collaboration
\cite{BaBar09}.
Instead of a pronounced growth of the scaled TFF with $Q^2$, observed
by \babar above 9~GeV$^2$, the Belle data are compatible (with the
exception of one point) at high $Q^2$ with the limit set by 
perturbative QCD \cite{BL81-2phot}:
$Q^2F^{\gamma^{*}\gamma\pi^0}=\sqrt{2}f_{\pi}\bm{\approx}0.185$ \textbf{GeV} with
$f_{\pi}=131$~MeV.

The behavior of the $\pi^0$ TFF with increasing $Q^2$ can be forecast
within calculable theoretical errors using the theoretical tools of
QCD---perturbative and nonperturbative.
Perturbative QCD governs evolution and supplies the means to calculate
radiative corrections, whereas nonperturbative QCD provides the
modeling concepts and techniques to determine the $\pi^0$ distribution amplitude (DA)---in
leading-twist two and higher twists---and treat the hadronic structure
of the nearly on-mass-shell photon.

In a recent paper \cite{SBMP12}, we have discussed what one should
expect for the outcome of a measurement of the $\pi^0$ TFF at high
$Q^2$ according to QCD.
We argued that the expected scaling behavior should follow---within
uncertainties---the interpolation formula of Brodsky--Lepage
\cite{BL89}:
$
  F^{\gamma^*\gamma \pi}(Q^2)
=
  \left(\sqrt{2}f_\pi\right)/\left(4\pi^2f_\pi^2 + Q^2\right)
$.
This phenomenological formula links the TFF at $Q^2=0$, fixed by the
axial anomaly, with the QCD asymptotic limit $\sqrt{2}f_\pi$.
While plausible and useful in practical terms, this formula is not
derived from QCD.
Hence, it is of paramount importance to calculate the TFF within
QCD in order to obtain an expression that can replicate the
appearance of stasis in the scaled TFF above some $Q^2$ value as a
result of switching on parton-photon interactions controlled by QCD.
Such a saturating behavior of the TFF can be obtained from a formalism
developed in a series of papers
\cite{BMS01,BMS02,BMS03,MS09,BMPS11} that combines
the dispersive approach of light-cone sum rules (LCSR)s
\cite{BBK89,Kho99} (see also\cite{SY99,ABOP10})
with QCD sum rules that employ nonlocal condensates \cite{MR86,BR91}.
The latter scheme is used to derive the pion DA, while the former
serves to properly accommodate the hadronic content of the
low-virtuality photon.
Then, the TFF is defined by the LCSR
\begin{eqnarray}
  Q^2 F^{\gamma^*\gamma\pi}\left(Q^2\right)
= \!\!\!\!\!
&&
  \frac{\sqrt{2}}{3}f_\pi
    \left[
          \frac{Q^2}{m_{\rho}^2}
          \int_{x_{0}}^{1}
                          \exp\left(
                                    \frac{m_{\rho}^2-Q^2\bar{x}/x}{M^2}
                              \right) \right.
\nonumber \\
&& \times \left.          \bar{\rho}(Q^2,x) \frac{dx}{x}
         + \int_{0}^{x_0}
                         \!\! \bar{\rho}(Q^2,x) \frac{dx}{\bar{x}}
    \right]\, ,
\label{eq:LCSR-FF}
\end{eqnarray}
where the spectral density is given by
$\bar{\rho}(Q^2,x)=(Q^2+s)\rho^\text{pert}(Q^2,s)$
with
$
\rho^\text{pert}(Q^2,s)
=
  (1/\pi) {\rm Im}F^{\gamma^*\gamma^*\pi}
  \left(Q^2,-s-i\varepsilon\right)
$,
and the abbreviations
$\bar{x}=1-x$,
$s =\bar{x}Q^2/x$,
$x_0 = Q^2/(Q^2+s_0)$
have been used.
The hadronic content of the quasi-real photon in the TFF is taken
care of by means of the first term in Eq.\ (\ref{eq:LCSR-FF}), whereas
the partonic pointlike interactions above $s>s_0$ are described by the
second term which is calculable order by order in QCD perturbation
theory.
Our computation is performed at the level of the NLO spectral density
by taking into account the correction pointed out in \cite{ABOP10}.
The other parameters have the following values \cite{BMS02}:
$s_0=1.5$~GeV$^2$,
$m_{\rho}=0.77$~GeV,
while the Borel parameter is
$M^2=M_\text{2-pt}^2/\langle{x}\rangle_{Q^2} < 1$~GeV$^2$
from the two-point QCD sum rule for the $\rho$-meson
with
$M_\text{2-pt}^2\in[0.5 \div 0.8]$~GeV$^2$
(see \cite{BMPS11} for more details).

The twist-two pion DA in the formalism of Bakulev, Mikhailov,
and Stefanis (BMS) \cite{BMS01} is modeled in terms of two Gegenbauer
coefficients $a_2$ and $a_4$:
$
 \varphi_{\pi}^{(2)\rm BMS}(x)
=
 \varphi^{\rm asy}(x)
 \left[
       1 + a_2 C_{2}^{3/2}(2x-1) + a_4 C_{2}^{3/2}(2x-1)
 \right]
$,
where
$
 \varphi^{\rm asy}(x)
=
 6x\bar{x}
$
denotes the asymptotic (asy) $\pi^0$ DA.
The values of
$a_2(\mu^2)=0.20$ and $a_4(\mu^2)=-0.14$ (at $\mu^2=1$~GeV$^2$)
are selected in such a way that the first ten moments
$
\langle \xi^{N} \rangle_{\pi}
\equiv
  \int_{0}^{1} dx (2x-1)^{N} \varphi_{\pi}^{(2)}(x,\mu^2)
$
with the normalization condition
$\int_{0}^{1} dx \varphi_{\pi}^{(2)}(x, \mu^2)=1$
lie inside a particular range determined in \cite{BMS01},
while all higher coefficients $a_6, a_8, a_{10}$ were determined and
found to be negligible.
This procedure gives rise to a whole ``bunch'' of possible DAs whose
shape is characterized by a double-humped structure with strongly
suppressed endpoints $x=0,1$.
This suppression is controlled by the vacuum quark virtuality
$\lambda_{q}^{2}(\mu^2=1~{\rm GeV}^2) \approx 0.4$~GeV$^2$ which
characterizes the nonlocality of the quark (quark-gluon)
condensate \cite{BMS01} and corresponds to a correlation length
of about 0.31~fm.
This parameter turns out to be intimately related to the Twist-four
coupling $\delta^2(\mu^2)$ which has a value around
$\delta^2 \approx \lambda_{q}^{2}/2$---details can be found in
\cite{BMS02}.
The \babar data from $Q^{2}=10$~GeV$^2$ onward show a marked
tendency to increase with $Q^2$ and are therefore incompatible with
the BMS formalism.
These data can be best reproduced with a flat-top $\pi^0$ DA
\cite{Dor09,Rad09,Pol09} that yields an auxetic TFF
\cite{SBMP12}---more below.

\begin{figure}[ht!]
\centerline{\includegraphics[width=0.48\textwidth]{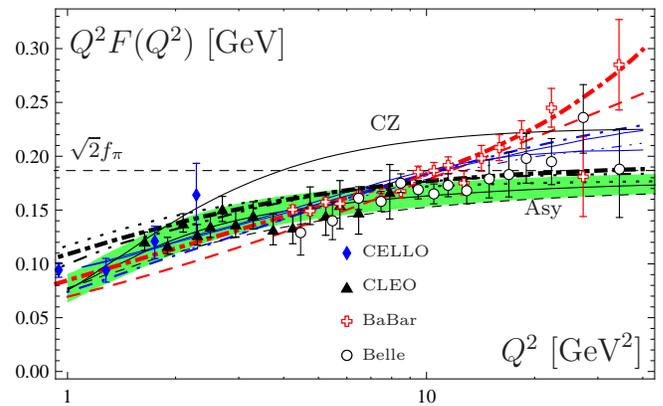}}
  \caption{\label{fig:pionFF-all} (color online).
    Theoretical predictions for the scaled
    $\gamma^*\gamma\pi^0$
    TFF, calculated in various approaches, in comparison with all
    existing experimental data, using for
    the latter the indicated notations.
    The shaded (green) band contains the results obtained
    within the BMS formalism in \protect\cite{BMPS11}, with the solid
    line inside it denoting the prediction for the BMS model DA
    \protect\cite{BMS01}.
    The other curves are explained in the text.}
\end{figure}

Let us look more closely in Fig.\ \ref{fig:pionFF-all} at what the
experimental data from different collaborations CELLO \cite{CELLO91},
CLEO \cite{CLEO98}, \babar \cite{BaBar09}, and Belle \cite{Belle12}
mean, comparing them with the results of several theoretical 
approaches.
The shaded (green) band shows the predictions calculated using the BMS
formalism detailed in \cite{BMPS11}.
The result for the BMS model \cite{BMS01} is represented by the solid
line inside it, while the width of the band collects uncertainties
from different sources:
(i) the variation of the shape of the $\pi^0$ DA extracted from QCD
sum rules with nonlocal condensates \cite{BMS01},
(ii) the uncertainty of the Twist-four coupling $\delta^2$
\cite{BMS02}, and
(iii) the sum of the Twist-six term \cite{ABOP10} and the
next-to-next-to-leading order (NNLO)
radiative correction, proportional to the $\beta_0$ function,
computed in \cite{MMP02}.
The combined treatment of these last two uncertainties is justified
because for the Borel parameter $M^2\leq 1$~GeV$^2$, we are
adopting, see, e.g., \cite{Kho99}, both have rather small magnitudes
comparable in size but opposite signs, with the Twist-six term being
positive.

Dropping extraneous details, let us address the other curves shown
in this figure from bottom to top.
The dashed line at the lower border of the shaded (green) band---tagged
Asy---denotes the TFF computed with the asymptotic $\pi^0$ DA.
Next, the dotted \cite{BCT11} and the double-dotted-dashed \cite{GR08}
lines, partly intersecting with the band, give the results obtained
with two holographic models based on the AdS/CFT correspondence.
The thick dashed-dotted line at the upper boundary of the shaded band
shows the recent prediction for the TFF obtained in \cite{CIKS12} using
an extended vector-dominance model.
The intermediary (blue) solid lines denote the TFF obtained for models
III (lower curve) and I (higher curve) from the LCSR analysis in
\cite{ABOP10}, whereas the prediction of their model II lies in
between (not shown).
The key characteristic of these models for the $\pi^0$ DA is
the large size of the coefficient $a_4>a_2$ and the non-negligible
values of the higher Gegenbauer coefficients
$a_6, a_8, a_{10}, a_{12}$.
Actually, model I is a flat-top DA with a reduced coefficient
$a_{2}^{\rm flat-top}(\mu^{2}=1~{\rm GeV}^2)\to 0.130$, while model III
has the coefficients $a_2=0.160, a_4=0.220, a_6=0.080$ (all values
at 1~GeV$^2$) \cite{ABOP10}---neglecting higher ones.
Note that this analysis takes into account the twist-six
contribution---found to be positive and very small for the
adopted value of the
Borel parameter $M^2=1.5$~GeV$^2$, while it ignores the explicit
inclusion of the (negative) NNLO radiative correction.
As one sees from this figure, the inverse hierarchy of the Gegenbauer
coefficients $a_{2} < a_{4}$ in the LCSR approach of \cite{ABOP10}
turns out to be not really enough to fully account for the auxetic
behavior of the \babar data above 9~GeV$^2$, while for the same
reason it causes a deviation from the Belle data and the asymptotic
QCD limit.

A similar result---shown as a double-dotted-dashed (blue) line
(Fig.\ \ref{fig:pionFF-all}) in the vicinity of the previous
two curves---was obtained by Kroll in \cite{Kro10sud} using a different
theoretical framework based on the Botts--Li--Sterman Modified
Factorization Scheme \cite{BS89,LS92}.
At the upper end of this regime of predictions one finds the TFF
(long-dashed-dotted line), computed by Polyakov in \cite{Pol09} by
using a flat-like $\pi^0$ DA, that was extracted from an effective
chiral quark model based on the instanton vacuum of QCD and including
leading-order (LO) evolution effects.
The two (red) curves shown at the top represent the predictions
obtained with a constant flat-top DA by Radyushkin \cite{Rad09}
(dashed line) and Lih and Geng \cite{LG12} (thick dashed-dotted line).
For completion, also the prediction for the TFF for the
Chernyak-Zhitnitsky (CZ) $\pi^0$ DA \cite{CZ84} is shown
(solitary line with the CZ label).

The main message from Fig.\ \ref{fig:pionFF-all} is that only with
a flat-top pion DA one can really replicate the auxetic behavior of
the $\pi^0\to\gamma^*\gamma$ transition form factor exhibited by the
high-$Q^2$ \babar data \cite{Dor09,Rad09,Pol09}.
It is worth noting, however, that to get best agreement in the
statistical sense, one has to ``cull'' the two outliers
at 14 and 27~GeV$^2$ before analysis.
These two outliers are close to the asymptotic QCD limit and
hence, strictly speaking, incompatible with a flat-top $\pi^0$ DA.
From the theoretical point of view, a flat-top DA reflects a
contingency approach geared in hindsight with the only aim to emulate
the auxetic trend of the \babar data for the pion TFF.
There are no features in some region of the longitudinal momentum
fraction $x$ of the valence quark and no humps anywhere.
Of course, this scaling must eventually break down at the endpoints
$x=0$ and $x=1$ because no parton can carry a larger momentum fraction
than 1.
Therefore, a flat-top $\pi^0$ DA describes the pion as being a
``pointlike'' particle with no internal structure because it
looks the same everywhere \cite{RRBGGT10}.
On the other hand, the BMS formalism yields predictions that
are possible only if the $\pi^0$ TFF is governed by QCD---in lingo:
collinear factorization underlying Eq.\ (\ref{eq:LCSR-FF}).
Accordingly, the TFF saturates at high momentum values of $Q^2$,
meaning that the high-virtuality photon couples to a single parton
inside the considered meson which is well-described by a BMS-like DA
that incorporates the nonlocality of the QCD vacuum.
In contrast, the \babar data \cite{BaBar09} for the pion contradict
this behavior necessitating some nonperturbative mechanism capable
of providing, e.g., $\ln(Q^2/\mu^2)$ enhancement
(where $\mu^2$ is some contextual nonperturbative scale $< 1$~GeV$^2$)
to the TFF, as encoded in a flat-top $\pi^0$ DA
\cite{Dor09,Rad09,Pol09}.
In short, the invention of a flat-like DA within a specific context
is contingent on the unforeseen behavior of the $\pi^0$ \babar data
at high $Q^2$.
Would the outcome of these data be in line with the asymptotic
QCD prediction, attempts to create a flat-like $\pi^0$ DA to challenge
the collinear QCD factorization would appear contrived and artificial.
Relying instead solely on the Belle data \cite{Belle12}, there would be
no need at all to invoke a flat-like DA based explanation of the
$\pi^0$ TFF.
\begin{figure}[t!]
 \centerline{\hspace{0mm}\includegraphics[width=0.48\textwidth]{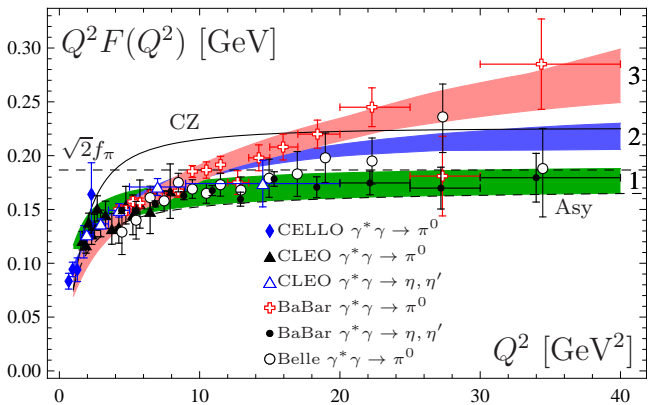}}
  \caption{\label{fig:strips} (color online).
    Classification of theoretical predictions for the spacelike
    scaled TFFs of the $\pi^0$ and the nonstrange component
    $|n\rangle$ of the $\eta$, and $\eta'$ in terms of three
    distinct shaded bands in comparison with all existing
    experimental data with notations as indicated.
    The predictions for the asymptotic (lowest dashed line) and the
    CZ DAs (solid line) are explicitly labeled.
    }
\end{figure}

The dichotomy between scaling of the TFF with $Q^2$ and auxesis becomes
more evident by inspecting Fig.\ \ref{fig:strips} which summarizes the
experimental situation for the $\pi^0$ TFF in conjunction with the
predictions of various theoretical approaches.
The latter can be organized into three distinct shaded bands, whose
widths are adjusted to include similar predictions obtained in
different theoretical schemes and are not related to particular
inherent theoretical uncertainties (like the BMS band shown in
Fig.\ \ref{fig:pionFF-all}).
The main goal of presenting Fig.\ \ref{fig:strips} is to place the
$\pi^0$ TFF in a broader perspective by collecting and comparing what
is known experimentally and categorizing what has been
proposed theoretically.
The central observation is that the experimental data for the
$\pi^0$ TFF arrange themselves at high-$Q^2$ values (starting at
10~GeV$^2$) along two branches which below this value unify into a
single one.
The upper branch of the data consists only of the \babar data for the
$\pi^0$ TFF \cite{BaBar09}---with the exception of the two outliers
downwards at 14 and 27~GeV$^2$, already mentioned in the
discussion of Fig.\ \ref{fig:pionFF-all}, and the Belle outlier
upwards at $Q^{2}=27.33$~GeV$^2$.
The lower branch contains the two \babar outliers and all the other
Belle data.
One can also count to this branch the
data for $(3/5)Q^2F^{\gamma^{*}\gamma n}$, where
$|n\rangle =(1/2)\left( |u\bar{u}\rangle + |d\bar{d}\rangle \right)$,
extracted from the \babar data \cite{BaBar11-BMS} for $\eta$ and
$\eta'$---see \cite{SBMP12} for more.
\begin{table*}[th!]
\begin{center}
\begin{ruledtabular}
\caption{
 We show in the first row the $\chi^2/\text{ndf}$ for the BMS model
 \cite{BMS01} in comparison with estimates of the coefficients
 $a_n$ of the $\pi$ DA determined by fitting the pion-photon TFF within
 LCSRs.
 The second and third rows show a 2D fit in the $(a_2,a_4)$ plane,
 while the last two rows employ a nonzero coefficient $a_6$.
 The errors are due to statistical uncertainties and a systematic error
 related to the Twist-four term.
 The last column shows the values of $\chi^2/\text{ndf}$
 (with ndf~$=$~number of degrees of freedom) for the considered data 
 sets.
 [All entries evaluated at
 $\mu_\text{SY}^2=5.76$~GeV$^2$ with SY abbreviating Schmedding
 and Yakovlev \cite{SY99}].
 \label{tab:fit.results}}
\smallskip
\begin{tabular}{lcccc}
 Data set                          &$a_2(\mu_\text{SY}^2)$ &$a_4(\mu_\text{SY}^2)$  &$a_6(\mu_\text{SY}^2)$ &$\chi^2/\text{ndf}$\\ \hline 
 CELLO \cite{CELLO91},
 CLEO \cite{CLEO98},
 Belle \cite{Belle12}              &$0.142$                &$-0.090$                &$0$                    &22.1/33\\ \hline
 CELLO, CLEO, Belle                &$0.154\pm0.046\pm0.055$&$-0.066\pm0.067\pm0.058$&$0$                    &20.1/31\\
 CELLO, CLEO,
 \babar \cite{BaBar09}             &$0.090\pm0.037\pm0.050$&$ ~~0.069\pm0.057\pm0.053$&$0$                  &69.5/33\\
 CELLO, CLEO, Belle                &$0.157\pm0.057\pm0.056$&$-0.192\pm0.122\pm0.077$&$0.226\pm0.161\pm0.033$&13.1/30\\
 CELLO, CLEO, \babar               &$0.177\pm0.054\pm0.056$&$-0.171\pm0.103\pm0.071$&$0.307\pm0.096\pm0.024$&33.3/32  
 \end{tabular}
\end{ruledtabular}
\end{center}\vspace*{-3mm}
\end{table*}

The classification of the theoretical predictions follows roughly this
pattern plus an additional band in between.
Those predictions agreeing with the standard QCD factorization scheme
are forming the lower (dark-green) band, labeled 1.
To be specific, this band contains the predictions obtained within the
BMS formalism in our most recent analysis in
\cite{BMPS11,BMPS_QCD2011,Ste_LC2011,SBMP11}, where one can find the
details.
Moreover, band 1 includes the result of the form-factor modeling of
\cite{CIKS12}, based on an extended vector-dominance model.
Interestingly, also the predictions from two different AdS/QCD models,
viz., \cite{GR08} and \cite{BCT11,BCT11b}, lie within band 1.
The lower boundary of this band is compatible with the asymptotic
$\pi^0$ DA (dashed line with the flag Asy).

The (red) band 3 collects the results from \cite{Rad09} and
\cite{LG12}, which both employ a flat-top $\pi^0$ DA, as well as
that of the analysis in \cite{LiMi09} which utilizes a flat-like DA
and Sudakov effects.
The predictions from \cite{KOT11b,KOT10plb} and \cite{MS12}, based on
a dispersive representation of the axial anomaly and quark-hadron
duality, are within this band as well.
The similar results of \cite{PP11} are also incorporated, while
the findings of \cite{WH10,WH11} (not shown) would appear just below
the edge of band 3.
Moreover, band 3 contains the result of the calculation in
\cite{McKPR11} which ascribes the auxetic TFF behavior to new physics
in the $\tau$ sector.

Between the two aforementioned bands, one has another class of
theoretical predictions forming the (blue) band 2.
This consists of the results obtained with models I, II, and III of
the LCSR analysis of \cite{ABOP10}, Polyakov's result \cite{Pol09},
extracted from the chiral quark model, and also an analogous result
from \cite{NV10}.
Note that the curve representing the TFF computed with the CZ DA
(shown as a single solid line with the label CZ)
joins band 2 at the far end of the experimentally accessible momentum
region, while it is more than $4\sigma$ away from all data below 
10~GeV$^2$.
Except the calculations giving rise to strip 3, most other predictions
shown in Fig.\ \ref{fig:strips} include
Efremov-Radyushkin-Brodsky-Lepage (ERBL) \cite{ER80,LB80} evolution
at the LO or NLO level.

In Table \ref{tab:fit.results}, we present statistical fits
of the pion TFF, calculated with LCSRs to the data.
The first row shows the original BMS values from \cite{BMS01},
while the next two rows show a 2D fit to two sets of data
using only the first two coefficients $a_2$ and $a_4$.
One set contains the CELLO, CLEO and Belle data, while the other one
consists of the CELLO, CLEO and \babar data.
The last two rows represent an analogous fit which also employs the
next coefficient $a_6$.
Evidently, the Belle data allow in both cases a better statistical
description.
This distinct behavior completely agrees with our classification scheme
presented in Fig.\ \ref{fig:strips}, with the BMS model and the
2D best-fit being entirely inside band 1.

\begin{figure}[h!]
 \centerline{\includegraphics[width=0.48\textwidth]{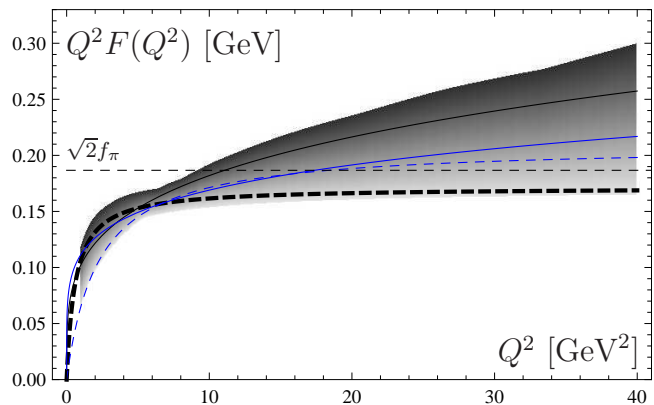}}
\caption{(color online).
    Unified range of theoretical predictions (grey area) discussed
    in context with Fig.\ \ref{fig:strips}.
    The darker shading indicates transition from scaling with $Q^2$ to
    auxesis.
    The experimental data of CLEO, Belle, and \babar are shown in terms of fits
    (single lines) explained in the text.
\protect\label{fig:grey-area}
           }
\end{figure}

But from another more data-oriented point of view, one may
argue (see, e.g., \cite{NV12}) that one should not distinguish the data
as we did in Fig.\ \ref{fig:strips} because the relative deviation
between the Belle and the \babar fits does not exceed $1.5\sigma$
\cite{Belle12}.
Appealing to this comparability, one may be tempted to unify the
theoretical results, shown in terms of three distinct strips in the
previous figure, into a single wide band.
This is illustrated in Fig.\ \ref{fig:grey-area}, where the darker 
shading of the band towards the top indicates the deviation from 
scaling with $Q^2$ predicted by QCD towards an auxetic TFF behavior 
contingent on model-dependent explanations.
Thus, even if such data pooling seems statistically acceptable, the
underlying theoretical approaches are hardly comparable to each other,
so that an interpretation of the Belle and the \babar data in relation
to each other and against some common standard appears rather 
questionable.
In this figure, all data sets are represented by single fits as 
follows:
(i) The top solid line shows a power-law fit
$Q^2|F(Q^2)|=A(Q^2/10~{\rm GeV}^2)^\beta$ to the \babar data with
$A=0.182,~\beta=0.25$ \cite{BaBar09}.
(ii) The lower (blue) solid line is also a power-law fit to the Belle 
data with $A=0.169$~GeV,~$\beta=0.18$ \cite{Belle12}, whereas the
dashed (blue) line below denotes a dipole fit 
$Q^2|F(Q^2)|=BQ^2/(Q^2+C)$ to the
Belle data with $B=0.209$~GeV,~$C=2.2$~GeV$^2$ \cite{Belle12}.
(iii) The thick dashed line at the bottom shows the dipole fit to the 
CLEO data \cite{CLEO98} with $B=0.171$~GeV,~$C=0.6$~GeV$^2$.

The appearance of the Belle data \cite{Belle12} on the
$\pi^0$ TFF forces us to consider the two-photon processes of light
pseudoscalar mesons in a greater perspective because the trend of
these data shows an antithetic behavior relative to that reported
by \babar \cite{BaBar09} with a relative deviation of about
$1.5\sigma$ \cite{Belle12}.
Instead of a pronounced rise with $Q^2$, it levels off and follows
more or less the scaling behavior predicted by QCD and collinear
factorization.
Though this is welcome from the theoretical point of view, an
increasing TFF behavior with $Q^2$ cannot be ruled out at present---at
least as long as no possible sources of errors have been identified by
the \babar Collaboration to revoke the validity of their results.
However, given that no unique QCD mechanism has been proposed to
provide the necessary enhancement of the $\pi^0$ TFF in order to
reconcile the \babar data with QCD, it seems reasonable to consider the
auxetic behavior of the \babar data and the entailed discrepancy to QCD
as an anomaly.
This view is strengthened by the fact that the pivotal QCD effects,
notably, the NLO and (the main) NNLO radiative corrections, ERBL
evolution, and the Twist-four term give suppression to the TFF,
except the Twist-six correction which is either small or of the same
size as the NNLO term but with opposite sign, hence canceling against
it.
Moreover, including in the theoretical analysis a small but finite
virtuality of the quasi-real photon, as in real single-tagged
experiments, also yields to suppression
\cite{ArBr10,Lic10,CIKS12,SBMP12}, albeit this suppression is more
significant at lower $Q^2$ values.

Bottom line is that skepticism about the accuracy of the \babar
data at high $Q^2$ prevails, giving preference to the Belle results
that found no deviation from the standard QCD scheme.
Moreover, in that case there is no chasm between the TFFs of the
$\pi^0$ and the $|n\rangle$, i.e., no evidence for a significant
flavor-symmetry breaking in the pseudoscalar meson sector of QCD---in
accordance with all previous experiments.

\acknowledgments
Two of us (A.P.B. and A.V.P.) are thankful to Prof. E. Epelbaum
and Prof. M. Polyakov for the warm hospitality at Bochum
University, where most of this investigation was carried out.
This work was supported in part by the Heisenberg--Landau Program under
Grant 2012, the Russian Foundation for Fundamental Research
(Grant No.\ 12-02-00613a), and the BRFBR--JINR Cooperation Program under
contract No.\ F10D-002.
The work of A.V.P. was supported in part
by HadronPhysics2, Spanish Ministerio de Economia y Competitividad and EU
FEDER under contract FPA2010-21750-C02-01, AIC10-D-000598, and
GVPrometeo2009/129.


\newcommand{\noopsort}[1]{} \newcommand{\printfirst}[2]{#1}
\newcommand{\singleletter}[1]{#1} \newcommand{\switchargs}[2]{#2#1}

\end{document}